\documentclass[12pt]{article}
\usepackage{graphicx}
\usepackage{dcolumn}
\usepackage{bm}

\begin{document}
\title{\bf Static and Dry Friction due to Multiscale Surface Roughness}
\vskip 1cm
\author{ J. B. Sokoloff\\ Department of Physics and\\Center for Interdisciplinary Research on Complex Systems\\
 Northeastern University, Boston, Massachusetts 02115, U.S.A., \\E-mail address: j.sokoloff@neu.edu}
\maketitle


\begin{abstract}
It is shown on the basis of scaling arguments that a disordered interface 
between two elastic solids will quite generally exhibit static and "dry friction" 
(i.e., kinetic friction which does not vanish as the sliding velocity approaches zero), because of 
Tomlinson model instabilities that occur for small length scale asperities. This provides a possible 
explanation for why static and "dry" friction are virtually always observed, and superlubricity almost never 
occurs.
\end{abstract}


\maketitle

\section{Introduction}

A few years ago, Muser and Robbins\cite{muser} proposed that static friction between clean crystalline solid surfaces, whose crystal axes were rotated by an arbitrary angle with respect to each other so that the surfaces are incommensurate, would be zero. This is commonly known as "superlubricity"\cite{superlub}. In order to explain the fact that superlubricity is rarely observed, they then proposed that the likely source of static friction for such surfaces is the existence of mobile dirt molecules, always present at any interface. The mobile molecules seek out local energy minima at the interface, which results in the surfaces being pinned with respect to each other, i.e., there is static friction. Actually, there can also be superlubricity for smooth disordered surfaces. In Refs. \cite{sokoloff3,lub} 
it was argued that a flat disordered interface between two 
macroscopic size surfaces which do not interact chemically will exhibit 
effectively no static friction for interface interaction per unit interface area
small compared to the shear elastic constant and high friction 
for interface interaction above this value. 
This frictionless regime is known as the weak 
pinning regime and the large friction regime is known as the strong 
pinning regime. In Ref. \cite{sokoloff3}, I argued that the existence of micron length scale asperities at an interface can result in both slow speed kinetic (often referred to as "dry friction") and static friction, resulting from the fact that these asperities can occur in multistable equilibrium configurations, which result in "dry friction" as a consequence of the Tomlinson model\cite{caroli} and static friction as well. The requirement for asperities being multistable is that the force on an asperity on one surface due to its interaction with a second surface dominate over the elastic forces resulting from its distortion, as a consequence of its interaction with the second surface. Since the interaction of an asperity with the second surface varies on atomic length scales, the distance over which the asperity distorts is negligibly small compared with its size (about a micron). Since micron length scales are still large compared with atomic spacings, however, the resulting static and dry friction is still likely to be quite small, indicating that the Muser-Robbins picture is still basically correct.  

It is argued here that sub-micron scale roughness 
can have a crucial effect on static friction between 
two three dimensional elastic solids that interact at their interface, 
primarily through inter-atomic hard core interaction, which will be 
the case for sufficiently high load \cite {izhak}. The basic idea is that the multiscale nature of the surface roughness results in the load being supported by a very small percentage of the atoms that would be in contact with the second surface, if the surface were perfectly flat. Consequently, the forces acting on these atoms by the second surface will dominate over the elastic restoring forces. Therefore, the shorter length scale asperities will be quite likely to exhibit multistability. This mechanism was previously proposed as a possible mechanism for explaining the ability of hard surface coatings to function as good lubricants\cite{carbon}. 
Plasticity (which was not discussed in Ref. \cite{carbon} but is discussed here) will likely play an important role for small length scale asperities as well. It will be argued that plastic deformation under the high degree of stress acting on the small length scale asperities will enhance the tendency for multistability to occur for these asperities, resulting in an increased tendency for the occurrence of static and "dry" friction. Recent work on contact mechanics\cite{persson3,hyun} make it possible to provide a more complete treatment of the scaling theory of Ref. \cite{carbon}, allowing a more numerical study of the conditions under which static friction can arise as a consequence of multiscale roughness, without the need to postulate that static and dry friction can only occur when there are mobile ion dirt particles present at the interface\cite{muser}, at least for surfaces that can be modeled as self-affine over a few length scales.

The next two sections summarize the results of Ref. \cite{carbon} and make some improvements and corrections. In section IV, the scaling theory of Ref. \cite{carbon} is expressed in the language of Persson's theory of contact mechanics\cite{persson3}, and the results of this theory are used to obtain information on the conditions under which large static and "dry" friction are likely to occur.

\section{Static Friction due to Hard Core interactions}  

As in Ref. \cite{carbon}, we consider two surfaces in contact which are disordered so that 
those atoms which are in contact are randomly 
distributed over the interface. In order to simplify the discussion, we will consider as our model for the interface, a rough elastic surface in contact with a perfectly atomically flat rigid surface. This has been argued in the past to be equivalent to the more correct model of two rough elastic surfaces in contact, for the contact mechanics problem, in which there are only forces normal to the surface\cite{johnson,persson3,hyun}. While it is not a rigorously correct representation of the problem when there are frictional (i.e., shear) stresses present as well, it should give correct orders of magnitude for the  scaling treatment considered on Ref. \cite{carbon} and here. Let us also assume that the atoms in contact at the interface interact only with hard 
core interactions. This could occur either because the surface atoms 
are chemically inert and there is negligible adhesion, or because the 
surfaces are being pushed together with a sufficient load so that 
the hard core interactions dominate. Let $\sigma$ denote the load 
or normal force per unit interface area, a, the mean atomic 
spacing and c, the fraction of the surface atoms of one 
surface that are in contact with the second surface. Then, each of the 
atoms in contact must contribute on the average to the normal force 
a force of order $\sigma a^2/c$. Since the force due to the hard core 
interaction between a pair of atoms acts along the line joining 
the atoms, for most relative positions of the atoms, it has a 
component along the interface, as illustrated in Fig. 1. 

\begin{figure}
\center{\includegraphics [angle=0,width=8cm]{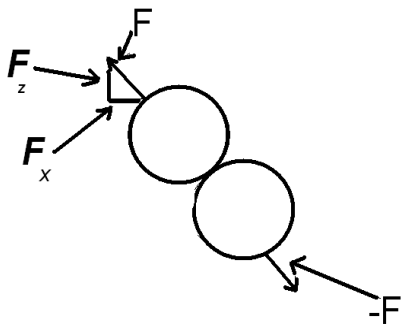}}
\caption{This figure illustrates how the hard core interaction between 
a pair of atoms, one belonging to each of the surfaces in contact, can 
both support the load and give rise to static friction between the 
surfaces. Since the force F between the pair of atoms can have both 
a component normal to the interface, $F_z$, which contributes to the 
normal force supporting the load, and a component along the interface 
$F_x$, the mean value of $F_x$ must be proportional to the mean value 
of $F_z$.}
\label{fig4}
\end{figure}

In the strong pinning regime, each surface  atom will sink 
into an interface potential 
minimum at the expense of the elastic forces holding it in place. 
Such a minimum will generally occur at an interstitial region on 
the second surface. If we now attempt to slide the surfaces relative 
to each other, each of the atoms in contact with the second surface 
will now, as it gets pulled out of its potential minimum, exert a 
component of its hard core interaction with the second surface parallel 
to the interface directed so as to oppose the attempted sliding motion.  
This is identified with the static friction. Since each of these 
atoms must also provide a component  $\sigma a^2/c$ normal to the interface 
on the average, it is clear the static friction is proportional to 
the load. This is the case because the contribution to the load and the static friction for 
each pair of atom in contact is provided by the same hard core force 
acting between the atoms. The proportionality constant $\mu_s$ is identified with the 
coefficient of static friction which is not too much smaller than 1. 
This accounts for Amonton's law without the need to assume 
that the friction is proportional to an ill-defined area of real contact. 
In the weak pinning limit, the component along the interface of the 
hard core force is random, and hence, for an infinite interface area 
and hence an infinite number of interface atoms, the components 
along the interface of the hard core forces cancel, resulting in 
effectively no static friction in the thermodynamic or macroscopic solid 
limit. 


\section {Effects of Roughness (i.e., Asperities on Several Length Scales)}

For two perfectly flat surfaces in contact pushed together with 
sufficiently weak load, each containing N surface 
atoms, the mean force of static friction will be the sum of the 
components of these hard core forces parallel to the surface, 
which have random magnitudes and directions. The net 
static friction, which is the sum of these components, is proportional to $N^{1/2}$ 
because they do not add together coherently. This is known as the weak 
pinning regime\cite {larkin,fukuyama,fisher}. It was shown in 
Ref. \cite {larkin,fukuyama,fisher,sokoloff3} that for sufficiently weak 
interaction across the interface (which will be true at low loads), the 
system will be in this regime. (Although mica is probably the only material with 
complete atomic scale flatness, bare mica surfaces actually exhibit relatively 
high friction because they interact with each other quite strongly.) 
At higher loads, the atoms from each 
surface can be pushed sufficiently far into the regions between atoms on 
the second surface (resulting in the atoms from the second surface being 
pushed apart), as this will minimize the repulsive energy. When we attempt to slide 
the surfaces relative to each other, there will be a net component of hard 
core repulsive forces along the surface opposing the sliding, i.e., static 
friction. In this case, the net force of static friction will be proportional 
to N. For models for surfaces with single length scale roughness, like the 
Greenwood-Williamson model \cite {GW}, since there are of the order of 
$10^8$ atoms at an interface between two micron size asperities, since $N^{1/2}$ 
is $10^4$, the static friction is a factor of $10^4$ smaller in the weak 
pinning than in the strong pinning regime. From Ref. \cite {lub}, 
we find that for flat surfaces, the load per unit area $\sigma$ below which the 
interface is in the weak pinning regime is given by $\sigma\approx c^{1/2}K$, 
where c is the fraction of surface atoms in contact with the substrate and K 
is the shear elastic constant. Since 
$K\approx 10^{11} N/m^2$ for most solid materials, unless c is extremely 
small, we find on the basis of this argument that any flat surfaces in 
contact that interact with only hard core interactions will be in the weak 
pinning regime.

Of course, no surfaces are perfectly smooth, and in fact, there can be 
roughness on several length scales. Let us assume then that there 
are $n_m$ orders of length scales, which we represent as follows: We divide 
the surface, which is defined to be at the $n_m$ length scale, somewhat arbitrarily into $M_{n_m-1}$ regions, which we define to be 
the asperities at the $n_m-1^{st}$ order length scale, a fraction $c_{n_m-1}$ of which are in contact 
with the second surface. An asperity at this length scale is considered to be in contact 
with the second surface if some part of it is in contact with the second surface. 
Let N represent the number of atomic size regions (i.e., 2-dimensional unit cells) contained in the projection of the rough surface onto a perfectly flat surface. In the discussion in this paragraph, the asperities 
are treated as rigid "hills" on the surface. Effects of distortion of the 
asperities are treated later in this section. As mentioned above, at each pair of asperities in 
"contact" the contact is likely to only occur at selected isolated regions, 
which we may refer to as $n_m-1^{st}$ order asperities. In fact, we will divide the surface of each $n_m-1^{st}$ order asperity 
into $M_{n_m-2}$ $n_m-2^{nd}$ order regions, somewhat arbitrarily (as we did for the $n_m-1^{st}$ order asperities), a fraction $c_{n_m-2}$ are 
in contact, again, contact meaning that there are high regions within this region which are in contact. The interface between 
a pair of $n_m-2$ order asperities can be divided up into $n_m-3$ order 
asperities as well, $M_{n_m-3}$ asperities, a fraction $c_{n_m-3}$ of which are 
in contact, etc., until we have done $n_m$ sub-levels of this sub-division. 
The surfaces can never be truly self-affine\cite {persson3}, 
however, because when we reach atomic dimensions at n=0 order, this sub-division into 
smaller and smaller length scales terminates. The 
area of contact of a $0^{th}$ level (i.e., smallest) asperity will 
contain of the order of $N_0={N\over M_0 c_0 M_1 c_1 ...M_{n_m-1} c_{n_m-1}}$ atoms, a fraction $c_{a}$ of which are in 
contact. It is these atoms at the zeroth  
(i.e., the final) order of asperities which support the load. 

If the $n^{th}$ order set of asperities are in the weak 
pinning regime, the static friction acting on it is reduced by a factor 
$(c_{n}M_{n})^{1/2}$, because by the above arguments, the static 
friction forces from these asperities act incoherently, and 
if the atoms at the interfaces of $0^{th}$ order mini-asperities 
in contact are in the weak pinning regime, the static friction 
is reduced from the strong pinning regime value (i.e., $\mu_s$ 
not too much smaller than 1) by a factor $(c_{a} N_0)^{1/2}$. 
This the mechanism 
was proposed as a possible way to explain why coatings of stiff 
materials are good lubricants\cite {harrison}.  

The arguments in the above paragraphs are only correct if each 
asperity is completely rigid, as we have assumed that the 
elastic forces which oppose displacements of the points of contact 
with the substrate resulting from the forces that the substrate 
exerts on them are due to the bulk solid's elasticity. Since each asperity has 
some height, however, it can distort so as to move its interface 
with the substrate closer to its potential minimum, even without 
distorting the bulk solid. Thus, whereas 
a particular substrate with completely rigid asperities might 
be in the weak pinning regime, if the asperities are able to 
distort by a sufficient amount, it might be in the strong pinning 
regime. This may explain why it is that although the estimates 
given earlier in this section indicate that most solids should 
be in the weak pinning regime, this is not consistent with 
the magnitudes of the friction coefficients that are observed 
for most solids. 

Let us now consider the distortions of the asperities that 
occur in response to the substrate potential. In our model, there exists a bunch of smallest asperities (which will 
be considered the lowest or zeroth level) which are in contact 
with the substrate. There are several groups of these that 
are assumed to be attached to a bunch of larger next or first 
order asperities. Groups of these first order asperities are 
then attached to larger asperities which are known as second 
order asperities. This hierarchy continues until we reach an 
"asperity" of width equal to that of the whole interface. This 
hierarchy of asperities is illustrated schematically in Fig. 2. Note that 
this figure is only a schematic representation of the problem; it is 
not being suggested here that the asperities have a rectalinear shape. 
Possible effects of asperity shape are discussed in appendix A.
\begin{figure}
\center{\includegraphics [angle=0,width=8cm]{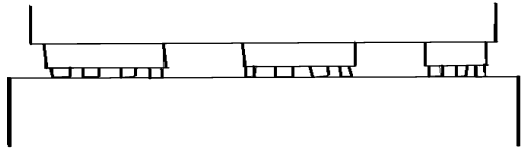}}
\caption{This is a schematic illustration of the asperity hierarchy 
on the top surface sliding on a flat substrate (i.e., the bottom block). 
(Real asperities have arbitrary shapes, as opposed to the square shapes 
shown in this schematic representation.)
Each asperity of a given order has a number of (smaller) asperities 
of one order lower on its surface. In turn, each of these lower order asperities 
has a number of (smaller) asperities of one order lower. This continues 
until we reach the zeroth order asperity, whose surface consists of 
atoms, although only three orders of asperities are illustrated here.}
\label{fig4}
\end{figure}

Consider the zeroth,  
the lowest order (i.e., the smallest), asperity. Let it 
have a height of order $L'_0$ and a width of order $L_0$. To find its 
distortion resulting from the sum of the substrate potential 
energies of all of the atoms of the asperity which are in 
contact with the substrate, we must minimize the sum of its elastic 
and substrate potential energies. The substrate potential energy 
is given by $V_0(c_a^{1/2} L_0/a) f_0(\Delta x_0/a)$, where $V_0$ is the 
amplitude of the interaction of a single atom with the substrate, 
resulting primarily from hard core repulsions between the atoms, 
$\Delta x_0$ is the amount that the surface in contact with the 
substrate slides under the influence of the substrate 
potential as the asperity distorts while all higher level 
asperities remain in an arbitrary rigid configuration and $f_0 (\Delta x_0/a)$ 
is a function of order unity which gives the variation of the 
substrate potential with $\Delta x_0$ for fixed, undistorted 
asperities of higher order (i.e., 
larger size in the present context). (Clearly, each of the 
zeroth order asperities has a different function; $f_0$ 
denotes a generic function describing the interface 
potential energy for a typical zeroth order asperity.) 
Each function clearly must 
possess multiple minima. We are assuming here that the 
surface of the asperity in contact with the substrate 
is in the weak pinning limit. The factor $(c_a^{1/2} L_0/a)$, which 
is of the order of the square root of the number of atoms 
in this surface expresses this fact. If the surface of the asperity 
in contact with the substrate is in the strong pinning limit instead, 
this factor will be replaced by $c_a (L_0/a)^2$, the number of 
atoms at the interface. [The factor of $c_a^{1/2}$ and $c_a$, 
respectively in these expressions were inadvertently not included 
in Ref. \cite{carbon}.] Treating this asperity as an elastic three 
dimensional solid in contact with the substrate we find from 
the discussion in Ref. \cite{lub} that the interface between the 
substrate and this asperity is in the weak pinning limit if 
$\sigma_0<c_a^{1/2}K$, where $\sigma_0$ is the mean load per unit 
interface area supported by this asperity and $c_a$ is the 
fraction of the surface atoms of this asperity that are 
in contact with the substrate. Assume that a fraction $c_0$ of 
the zeroth order asperities have atoms belonging to them 
in contact with the substrate. Let $c_1$ represent the 
fraction of next order (first order) asperities whose 
zeroth order asperities are in contact with the substrate, 
$c_2$, the fraction of second order asperities whose first 
order asperities have their zeroth order asperities in 
contact with the substrate, etc., up to $n_m^{th}$ order. 
Then $\sigma_0=\sigma/(c_0 c_1 c_2...c_{n_m-1})$, where $\sigma$ is the load per 
unit apparent area of the surface of the whole solid. 
Then, we conclude that the criterion for the atoms at 
the interface between the zeroth order asperity and 
the substrate to be in the weak pinning regime is that 
$\sigma<(c_a^{1/2}c_0 c_1 c_2...c_{n_m-1})K$. We see from this inequality 
that the more fractal the surface is, the more difficult 
it is for the zeroth order asperity to be in the weak 
pinning regime. The cost in elastic 
energy due to the shear distortion of the asperity 
can be determined by the following scaling argument: The 
elastic energy density for shear distortion of the asperity 
is proportional to $(\partial u_x/\partial z)^2,$ where $u_x$ 
represents the local displacement due to the distortion, the 
x-direction is along the interface and the z-direction is normal 
to it. The $u_x$ must scale with $\Delta x_0$ and the 
dependence of $u_x$ on z has a length scale $L'_0$. 
Thus the elastic strain energy of the asperity is 
of the order of $(1/2)L_0^2 L'_0 K(\Delta x_0/L'_0)^2$, where 
K is the shear elastic constant and $(\Delta x_0/L'_0)$ is 
the average shear strain and $L_0$ is the mean width of the asperity. 
Using a perhaps more correct non-rectangular shape for the asperities is shown in 
appendix A to only produce a correction of order unity. 
Minimizing the sum of these expressions 
for elastic and substrate potential energy, we obtain
$$\Delta x_0/a\approx (V_0/Ka^3)c_a^{1/2} (L'_0/L_0) f'_0 (\Delta x_0/a) \eqno (1)$$
since $f_0'$, the derivative of $f_0$ with respect to its argument, 
it is of order one, from the definition of $f_0$. Let us follow a 
line of reasoning like that of Ref. \cite{caroli}, a modified version 
of which is given in Ref. \cite{carbon}. For 
$(V_0/Ka^3)c_a^{1/2} (L'_0/L_0)$ below a certain value of order one, 
for small $V_0/Ka^3$, Eq. (1) 
can have only one solution for $\Delta x_0$. 
The reason for this is illustrated in Fig. 3. Under 
such circumstances the average kinetic friction, 
in the limit as the sliding velocity approaches zero, is zero. 
For a surface with 
an infinite number of asperities, distributed 
uniformly in space, it was shown in Ref. \cite{caroli} 
that the static friction is zero as well. A modified version 
of this argument, which points out that for a surface 
with a finite number of asperities the static friction 
is non-zero, but smaller by a factor of $a/L_0$ compared 
to what it would be if the contributions of the asperities 
to static friction acted coherently is provided in appendix B of Ref. \cite{carbon}. 
If this asperity is in the strong pinning limit instead, we 
replace the factor of $c_a^{1/2} (L_0/a)$ by $c_a (L_0/a)^2$ to account for this 
and as a result, the factor $c_a^{1/2} (L'_0/L_0)$ gets replaced by 
$c_a (L'_0/a)$, which could make the asperity 
satisfy the criterion for multistability more easily, and consequently, the 
friction from these asperities will no longer be reduced by the factor
$a/L_0$. For a load per unit area 
$\sigma$, assumed to be primarily due to hard core interactions, 
we may assume $V_0\approx \sigma a^3/c$, where c is the fraction 
of the surface atoms which are in contact with the substrate. By the 
above arguments, $c=c_a c_0 c_1 c_2...c_{n_m-1}$. Then, we see that the 
criterion for the zeroth order asperity to be multistable is 
$\sigma>c_0 c_1 c_2...c_{n_m-1} K$. If the 
criterion for weak pinning for the zeroth order asperity surface 
is not satisfied, the criterion for monostability of this asperity 
gets changed from the above inequality to $(V_0/Ka^3) c_a (L'_0/a)<1$, 
which is more difficult to satisfy since $L'_0/a$ can be considerably 
greater than 1.  

At the next level, we have an asperity surface in contact 
with the substrate which consists of a collection of 
the lowest level (i.e., the smallest) asperities 
discussed in the previous paragraph. Assuming this 
asperity to be in the weak pinning regime, the 
potential of interaction with the substrate, which is 
the sum of all of the interactions of the substrate 
with the lowest order asperities, which cover a first 
order asperity, is of order 
$V_0 (c_a c_0)^{1/2} (L_0/a)(L_1/L_0) f_1(\Delta x_1/a)$. Here $L_1$ 
represents the width of this order asperity, 
$\Delta x_1$ represents a displacement of the lower 
surface of this level asperity for fixed (i.e., un-distorted) 
configurations of all higher order asperities, 
and $f_1$ denotes one of the functions which describes 
the interface potential energy of one of the first order asperities. 
It has at least one minimum and 
runs over a range of magnitude one as its argument runs 
over a range of order one. [Again, the factor of $(c_a c_0)^{1/2}$ and $c_a c_0$, 
respectively in these expressions were inadvertently not included 
in Ref. \cite{carbon}.]The elastic energy is of the 
order of $(1/2)L'_1 L_1^2 K(\Delta x_1/L'_1)^2$, 
by the argument given above Eq. (1), where 
$L'_1$ is the height of the body of the first order 
asperity, which is assumed to be much larger than 
$L'_0$. Minimizing the sum of these two energies, we obtain 
$$\Delta x_1/a\approx (V_0/K a^3)(c_a c_0)^{1/2} (L'_1/L_1)f'_1 (\Delta x_1/a). \eqno (2)$$
Again, we conclude, based on the arguments presented in 
Ref. \cite {caroli}, 
that the static friction is reduced by a factor of $L_0/L_1$ below what 
it would be if the contributions to the static friction 
from each of the mini-asperities at this level acted coherently. 
If the zeroth order asperities attached to this first order 
asperity are in the strong pinning regime, the factor of 
$(c_a c_0)^{1/2} L_1/L_0$ in the equation for the interaction of this asperity  
with the substrate is replaced by $c_a c_0 (L_1/L_0)^2$, and hence, 
the right hand side of Eq. (2) has the factor $(c_a c_0)^{1/2} L'_1/L_1$ replaced 
by $c_a c_0 L'_1/L_0$, which can make the solutions to this equation 
for $\Delta x_1$ multistable. 
 
Continuing this procedure, we find that the displacement of the $n^{th}$ level 
mini-asperity is found by solving  
$$\Delta x_n/a\approx (V_0/Ka^3)(c_a ...c_{n-1})^{1/2} 
(L'_n/L_n) f'_n (\Delta x_n/a),\eqno (3)$$
where $L_n$ and $L'_n$ are the width and height of the 
body of the $n^{th}$  level asperity. [The factor of $(c_a ...c_{n-1})^{1/2}$, 
respectively in these expressions were inadvertently not included 
in Ref. \cite{carbon}. The general conclusions of Ref. \cite{carbon}, however, 
are not changed by this omission.]
If $(V_0/Ka^3)(c_a ...c_{n-1})^{1/2}=\sigma/[K(c_n c_{n+1}...c
_{n_m})^{1/2}]>1$, and $L'_n/L_n\sim 1$ for all n, 
asperities of all orders will be multistable, implying the occurrence 
of large static friction. If the condition given earlier for 
strong pinning at the zeroth order asperity interface, namely 
$\sigma>c_a^{1/2} c_0 c_1...c_{n_m} K=(c/c_a^{1/2})K$ is satisfied, 
the condition for multistability on all levels, $\sigma>(c_n c_{n+1}...c_{n_m})^{1/2} K$ 
is certainly satisfied for the zeroth order asperities, but may break down at some order n. 
Since more of the c's will appear in the product of the c's in this 
expression for lower order than higher order asperities, we conclude that the lower order 
asperities are more likely to be multistable than the higher order ones. If the $n^{th}$ 
order asperity is in the strong pinning regime, the factor $(c_a ...c_{n-1})^{1/2} 
(L'_n/L_n)$ in Eq. (3) gets replaced by $(c_a ...c_{n-1})(L'_n/a)$, which leads to the condition 
for multistability: $\sigma>(c_n c_{n+1}...c_{n_m})(L'_n/a)K.$

\begin{figure}
\center{\includegraphics [angle=0,width=8cm]{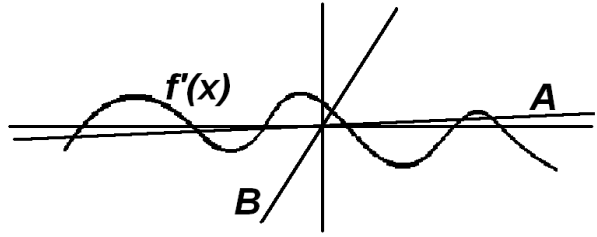}}
\caption{This figure illustrates the solution of Eqs. (1), (2) and 
(3), for $\Delta x_0$, $\Delta x_1$ and $\Delta x_n$, respectively. 
$f'(x)$ is a schematic illustration of the functions $f'_0$, $f'_1$ and 
$f'_n$, and x denotes $\Delta x_0$, $\Delta x_1$ or $\Delta x_n$, 
respectively. Lines A and B represent the line $y=(Ka^3/V_0)(L_n/L'_n)x,$ 
for $Ka^3/V_0)(L_n/L'_n)<1$ and $Ka^3/V_0)(L_n/L'_n)>1$, respectively. 
For the situation illustrated by line A, there are multiple solutions 
(i.e., multi-stability), while for the situation illustrated by line B, 
there is only one (mono-stability).}
\label{fig4}
\end{figure}

At least in the limit in which the stress acting on a given length scale asperity resulting from the load that it carries is small compared to Young's modulus, recent theories of contact mechanics for surfaces with multi-length-scale roughness\cite{persson3,hyun} show that the area of contact on all length scales (even the smallest) is approximately proportional to the applied load. This implies that although the smallest length scale asperities will certainly flatten out as the load increases (which means in the language of the present manuscript that a larger fraction of the subasperities on the surface of this asperity will come in contact with the second surface, and a similar picture holds for each of its subasperities). Consequently, on the basis of these theories, the amount of load carried by each of the atoms which are in contact with the second surface will not decrease as a consequence of more of these atoms being in contact with the second surface.
  
\section{Application of Persson's Contact Mechanics Theory}

Persson has developed an analytic contact mechanics theory for self-affine surfaces\cite{persson3}. In this section, a connection will be made between the parameters used here and the contact area on a given length scale $\lambda$, $A(\zeta)$ used in Ref. \cite{persson3}, were $\zeta=L/\lambda$, where L is the largest length scale of the surface. On a length scale $\lambda=L_{n_m}$, there will be $(L_{n_m}/L_{n_m-1})^2$ asperities of lateral size $L_{n_m-1}$, a fraction $c_{n_m-1}$ of which are in contact with the substrate. It is clear that $L_{n_m}$ denotes the largest length scale, which represents the  dimensions of the actual surface. On this surface, there will be $c_{n_m-2}(L_{n_m-1}/L_{n_m-2})^2$ contacting asperities of size $L_{n_m-2}$, and on each of these asperities, there will be $c_{n_m-3}(L_{n_m-2}/L_{n_m-3})^2$ contacting asperities of size $L_{n_m-3}$. As we continue, we reach the $L_{n+1}$ scale asperities, each of which contains $c_n (L_{n+1}/L_n)^2$ contacting asperities of linear dimension $L_n$. If we were to stop at this length scale and assume that the area of contact of each of these $L_n$ length scale asperities is smooth (because we are imagining that our "measuring instruments" cannot see such small length scales), and the total area of contact of the surface with the substrate will be given by the product of the above numbers of contacting asperities at each length scale and $L_n^2$ (the cross-sectional area of an $n^{th}$ scale asperity), or 
$$P(\zeta_n)=A(\zeta_n)/A(\zeta_{n_m})=$$
$$c_{n_m-1}(L_{n_m}/L_{n_m-1})^2 c_{n_m-2}(L_{n_m-1}/L_{n_m-2})^2$$ 
$$c_{n_m-3}(L_{n_m-2}/L_{n_m-3})^2 ... c_n (L_{n+1}/L_n)^2 ({L_n^2\over A})=$$
$$c_{n_m-1} c_{n_m-2} c_{n_m-3} ... c_{n+1} c_{n}, \eqno (4)$$
where we have used the fact that $A(L_{n_m})=L_{n_m}^2$, and hence, the quantity 
$P(\zeta_n)=A(L_n)/A(L_{n_m})$ is a quantity 
used in Persson's theory \cite{persson3}, where the magnification $\zeta_n=L_{n_m}/L_n$. The largest length scale $L_{n_m}$, which will often be denoted by L, is the length of sliding solid, and $A=L^2$ is its nominal area. Using this relationship between Persson's $P(\zeta_n)$ and the quantity $c_n$ used in Ref. \cite{carbon} and here, we can write the criterion for multistability of the $n^{th}$ order asperity discussed under Eq. (3) of the last section as 
$$\sigma>(a/L'_n)P(\zeta_n)K={a\over L'_n} {A(\zeta_n)\over A}K \eqno (5)$$
assuming strong pinning at all asperity interfaces and 
$$\sigma>(L_n/L'_n) [P(\zeta_n)P(\zeta_a)]^{1/2}K={L_n\over L'_n} {[A(\zeta_n)A(\zeta_a)]^{1/2}\over A}K, \eqno (6)$$
where $\zeta_a=L_{n_m}/a$, assuming weak pinning. From Eqs. (5) and (6), it is clear that for sufficiently small $A(\zeta_n)$, the asperities will tend to be multistable, resulting in nonzero static and dry friction. By writing the criteria for multistability of the asperities in terms of Persson's $P(\zeta_n)$, we can use the results of his theory of contact mechanics to determine the load dependence of these criteria. 

Let us now use Persson's theory to determine the contact area per asperity at a particular length scale (call it the $n^{th}$ length scale). Since the number of $n_m-1$ level asperities in contact is $c_{n_m-1}(L_{n_m}/L_{n_m-1})^2$, and the number of $n_m-2$ level asperities residing on one of the $n_m-1$ level asperities in contact is $c_{n_m-2}(L_{n_m-1}/L_{n_m-2})^2$, etc., we conclude that the number of $n^{th}$ level asperities in contact is 
$$c_{n_m-1}(L_{n_m}/L_{n_m-1})^2 c_{n_m-2}(L_{n_m-1}/L_{n_m-2})^2 c_{n_m-3}$$
$$(L_{n_m-2}/L_{n_m-3})^2...c_n (L_{n+1}/L_n)^2, \eqno (7)$$
which is equal to
$$c_{n_m-1}c_{n_m-2}c_{n_m-3}...c_n (L_{n_m}/L_n)^2=A(L_n)/L_n^2. \eqno (8)$$
Using the fact that the contact area per $n^{th}$ order asperity is equal to the nominal area of the interface $A(L_{n_m})$ divided by the above expression for the number of $n^{th}$ level asperities in contact, we obtain for the mean contact area per asperity at the $n^{th}$ level $L_n^2$, independent of the load. 
If we stop at $n^{th}$ order in Persson's treatment\cite{persson3}, we take the surfaces at smaller length scales to be perfectly flat. It is interesting to note that if we calculate the area per $n^{th}$ order asperity using Archard's theory of contact mechanics for fractal surfaces\cite{archard}, we find that it is proportional to the load to a power which decreases as n decreases, implying that even if the surface is assumed to be perfectly elastic, the smaller length scale asperities will flatten out at sufficiently high load, provided that we put a small length scale cut-off in his theory. In reality, there must also be a small enough length scale in Persson's theory as well in which asperities at all smaller length scales will get completely flattened out, if the solid is treated as being completely elastic, but of course long before that point, we must deal with plasticity. 

Persson shows using his theory of contact mechanics\cite{persson3} the not unexpected result that as the magnification parameter $\zeta_n=L_{n_m}/L_n$ approaches infinity $A_{el}(\zeta_n)/A$, where $A_{el}(\zeta_n)$ denotes the area of contact associated with elastic deformation of asperities, approaches zero. Then $E_Y A_{pl}(\zeta_n)=\sigma_0 A_0$, where $A_{pl}(\zeta_n)$ is the contact area associated with plastic deformation of asperities under load and $E_Y$ denotes the hardness of the material (i.e., the compressional stress at which plastic deformation takes place). Some of Persson's arguments are summarized and some steps are filled in appendix B. Generally, when lowest length scale asperities have failed plastically under compression, if one attempts to shear the asperity by sliding the surfaces relative to each other, it will also have failed plastically in shear motion\cite{spencer}, and hence, either the resulting shear stress due to shearing of the asperity will be nearly independent of the shear strain or it will increase with shear strain , but much more slowly, as a function of shear strain, than it would if its response were elastic. In the latter case, the tendency towards multistability will be greatly increased over what it would be if the shear response were elastic, implying an increase in the dry and static friction. In the former case, the asperity would shear without any increased restoring force, most likely indicating that it will break as it shears relative to the substrate. This indicates that the system has switched over to a regime in which a good part of the kinetic friction is due to wear (i.e., the breaking of the smallest length scale asperities). The static friction will likely go to zero if this were a completely correct model for plastic flow, because in such a case, there is no energy cost to shear the asperity.
 
As mentioned earlier, Persson has demonstrated that as $\zeta_n$ approaches infinity, all parts of the contact area exhibit plasticity.  (The argument is discussed in appendix B of this article). It is important, however, to estimate the value of $\zeta_n$ for which plastic deformation of asperities on the $n^{th}$ scale becomes important. The rough criterion that we will use to estimate this is when the mean value of the compressional stress over an $n^{th}$ order asperity becomes comparable to $E_Y$.  
One may determine when plasticity at the smaller length scale asperities sets in using Persson's theory, assuming that the distortions on all length scales are elastic until this point. Persson finds in the elastic regime the following expression for the contact 
area at $n^{th}$ scale magnification:
$${A(\zeta_n)\over A}=P(\zeta_n)={4\sigma (1-\nu^2)\over q_0 h_0 E} ({1-H\over \pi H})^{1/2} \zeta_n^{(H-1)}, \eqno (9)$$
where $H=3-D_f$, where $D_f$ is the fractal dimension of the surface, $q_0=2\pi/L$ and $h_0$ is the rms surface height fluctuation, and $\nu$ is Poisson's ratio. Since we must have 
$$<\sigma>|_{\zeta_n} A(\zeta_n)=\sigma A, \eqno (10)$$
$<\sigma>|_{\zeta_n}=\sigma/P(\zeta_n)$, and hence, from Eq. (9), we have
$${<\sigma>|_{\zeta_n}\over E}={q_0 h_0\over 4(1-\nu^2)} ({\pi H\over 1-H})^{1/2}\zeta_n^{(1-H)}. \eqno (11)$$
When $<\sigma>|_{\zeta_n}$ becomes comparable to $E_Y$, one must consider plasticity. Persson considers H=0.8, a value typical for pavement and $q_0 h_0$ equal to 0.001 and 0.01. [Smaller values of H represent rougher surfaces, because by the definition of a self-affine surface, reducing the length scale along the surface by a factor $\lambda$ (with $\lambda<1$) reduces the height fluctuations only by a factor $\lambda^H$, and since $H<1$, larger height fluctuations occur over smaller lateral length scales.] Using these numbers in the above expression for $<\sigma>|_{\zeta_n}$, we find that 
for $L/L_n=10^8$, typical for atomic level asperities, $<\sigma>|_{\zeta_n}/E\approx0.04$ for $q_0 h_0=0.001$ and 0.4 for $q_0 h_0=0.01$. 
$E_Y/E$=0.1 for glass, 0.0013 for steel and 0.0125 for hard plastic. Thus, atomic scale asperities are clearly in the plastic regime. For diamond films, $E_Y$, the hardness, is about 100 GPa\cite{lemoine}, and since E is 1080GPa\cite{kittel}, $E_Y/E\approx 0.09$. Other diamond-like carbon films with large $sp^3$ (i.e., tetrahedral $\sigma$) bond content also exhibit $E_Y/E$ of comparable magnitude. We see here the important role played by the relatively large values of $E_Y/E$ that exist for these materials.  On the basis of the above arguments, we conclude that it is possible for sufficiently smooth surfaces that for diamond-like carbon films, asperities on all length scales that exist for a crystalline material might be elastic. 

At this point, let us re-examine the argument presented in Ref. \cite{carbon}, that the large values of Young's modulus of stiff coatings, such as diamond-like carbon films, can still account for their excellent lubricating properties, making use of Eqs. (5), (6) and (9). Combining Eqs. (5) and (6) with Eq. (9), we obtain for the condition for the $n^{th}$ order asperities to be multistable 
$${K\over E}<{L'_n\over a} {q_0 h_0\over 4(1-\nu^2)} ({\pi H\over 1-H})^{1/2}\zeta_n^{(1-H)}, \eqno (12a)$$
if the interfaces between each level asperity less than or equal to $n^{th}$ order are in the strong pinning regime (defined in the previous two sections), or 
$${K\over E}<{L'_n\over L_n} {q_0 h_0\over 4(1-\nu^2)} ({\pi H\over 1-H})^{1/2}(\zeta_n\zeta_a)^{(1-H)\over 2}, \eqno (12b)$$
if the interfaces between each level asperity less than or equal to $n^{th}$ order are in the weak pinning regime. Since the n=0 order asperities are the ones most likely to be multistable, let us apply Eqs. (12a) and (12b) to the n=0 case, for which they reduce to 
$${K\over E}<{L'_0\over a} {q_0 h_0\over 4(1-\nu^2)} ({\pi H\over 1-H})^{1/2}\zeta_a^{(1-H)}, \eqno (13a)$$
and 
$${K\over E}<{L'_0\over L_0} {q_0 h_0\over 4(1-\nu^2)} ({\pi H\over 1-H})^{1/2}\zeta_a^{(1-H)}, \eqno (13b)$$
respectively. The condition for weak pinning at an n=0 asperity interface found in the last section is 
$$\sigma<c_a^{1/2}(c_0 c_1 ... c_{n_m-1})K, \eqno(14)$$
which when written in terms of Persson's notation\cite{persson3} is
$$\sigma<c_a^{-1/2}{A(\zeta_a)\over A}K.  \eqno (15)$$  
Substituting for $A(\zeta_a)/A$ using Eq. (9), this condition can be written as  
$${K\over E}>{c_a^{1/2} q_0 h_0\over 4(1-\nu^2)} ({\pi H\over 1-H})^{1/2}\zeta_a^{(1-H)}. \eqno (16)$$
In order for the zeroth order asperities to be multistable, and at the same time have their interfaces with the substrate in the weak pinning limit, we must have $c_a^{1/2}<L'_0/L_0$, which is not difficult to satisfy unless the zeroth order asperities are all extremely short and fat and/or have a large fraction of the atoms at their interfaces in contact with the second surface.

\section {Conclusions}

It has been shown, using a scaling theory of friction for surfaces with multiscale roughness\cite{carbon} combined with Persson's theory of contact mechanics, that, in contrast to surfaces with only single scale roughness which appear to always exhibit superlubricity, unless one postulates the existence of mobile dirt particles at the interface, surfaces with multiscale roughness will almost always exhibit static and dry friction, characteristic of almost all solid surfaces. The static and dry friction  come about because asperities at the smallest length scales will be multistable, because they support all of the load  if they remain elastic, or will exhibit plastic failure. It is argued that plastic failure will increase the tendency towards the asperities being multistable. Numerical finite element calculations used recently to study contact mechanics\cite{hyun} can also be applied to this problem. Since they appear to yield results for the contact mechanics problem which are qualitatively similar to those of Ref. \cite{persson3}, however, the results for the multistability (i.e. Tomlinson model\cite{caroli}) problem should not differ significantly from those presented here. There is one small difference, however, which is worth considering. Whereas Ref. \cite{persson3} finds that $A(\zeta_n)/A$ is a linear function of $\sigma$ in the elastic regime, Ref. \cite{hyun} reports that this function is a slightly sub-linear function of $\sigma$. As already suggested in Ref. \cite{carbon}, this would imply that as $\sigma$ increases, the interface could possibly switch from having all asperities in the monostable regime, resulting in low friction, to the multistable regime, resulting in high friction, when $\sigma$ becomes sufficiently large. 
 
\section {Appendix A: Effects of Asperity Shape on Friction}

In section III, we assumed that the actual shape of an asperity would have an insignificant effect on the multiscale Tomlinson model that has been developed here. In this section, we will examine whether this is likely to be a reasonable assumption. Asperities on all length scales are likely to initially be peaked, but as they are pressed together they flatten out. The amount of flattening of a given length scale asperity depends on the amount of load that asperity carries, which itself depends on the degree of compression of the asperity, as is known to happen for single length scale asperities\cite{johnson}. The amount of load carried by an asperity depends on the number of atoms in contact with the substrate, as we have seen earlier that each atom carries a load of the order of $Pa^2/c$. Typically, near the peak of an asperity at any length scale, the profile can be represented to a good approximation by a parabaloid whose cross-sectional area at a distance z from apex is proportional to z\cite{johnson}. It is not a bad approximation to assume that we can use this result for an asperity which gets compressed because it is in contact with the substrate. Then let us assume that a compressed asperity has a flat peak at a minimum value of z ($z_{min}$), whose area is proportional to $z_{min}$. Then, using the parabaloid approximation for the  asperity, we have $\pi r^2\approx 2\pi Rz$, let us imagine slicing each such asperity into slices of thickness $\Delta z$, each of whose area is proportional to z. Then, the shear stress between slices n and n+1 is equal to
$$(2\pi Rz_n/a)K[({u_{n+1}-u_n\over a})z_{n+1}-{u_{n}-u_{n-1}\over a})z_{n}], \eqno (1A)$$ 
following the method for taking the continuum limit of the harmonic approximation treatment of a discrete lattice discussed in Ref. \cite{ashcroft}. This must be zero when the asperity is in equilibrium. The continuum limit of this equation is 
$${\partial\over \partial z}({\partial (zu)\over\partial z})=0, \eqno (2A)$$
whose solution is 
$${\partial u\over\partial z}={c\over z} \eqno (3A)$$
where c is a constant. Since at $z=z_{min}$, we require that $K(\pi r_c^2)(\partial u/\partial z)|_{z=z_{min}}=\sigma\pi r_c^2$, where $r_c$ is the contact radius, and $\sigma$ is the shear stress acting on the contact area of the asperity, because this is the condition for equilibrium of the asperity. This gives us $c=\sigma z_{min}/K$. Then, $\partial u/\partial z=(\sigma z_{min}/Kz)$. We may find the elastic energy of an asperity from
$$E_{elast}=(1/2)\int_{z_{min}}^{z_{max}}dz 2\pi RzK({\partial u\over\partial z})^2=$$
$$[\pi R(\sigma z_{min})^2/K]\int_{z_{min}}^{z_{max}} {dz\over z}=[\pi R(\sigma z_{min})^2/K]
ln({z_{max}\over z_{min}}). \eqno (4A)$$
In order to compare this with the expression used in the scaling treatment of multiscale asperities used in Ref. \cite{carbon} and here, let us now write this expression in terms of $\Delta u$, given by
$$\Delta u=(\sigma z_{min}/K)\int_{z_{min}}^{z_{max}} {dz\over z}=(\sigma z_{min}/K)ln({z_{max}\over z_{min}}). \eqno (5A)$$
Solving for $\sigma$ in terms of $\Delta u$ and substituting in the above expression for $E_{elast}$, we obtain
$$E_{elast}={\pi KR\over ln({z_{max}\over z_{min}})}\Delta u^2={K\pi L^2z_{max}\over 2 ln({z_{max}\over z_{min}})}({\Delta u\over z_{max}})^2, \eqno (6A)$$
where L is defined by $\pi L^2=\pi r_{max}^2=2\pi R z_{max}$. For the $n^{th}$ order asperity, we identify L with $L_n$ and $z_{max}- z_{min}$ with $L'_n$. Since for all except for the extremes of complete compression and nonexistent compression, $z_{max}$ and $z_{max}-z_{min}$ are of comparable magnitudes and $ln(z_{max}/z_{min})$ contributes only a factor of order unity, we conclude that our original scaling treatment of multiscale asperity multistability, which does not consider the actual shape of the asperities, will give correct orders of magnitude.  

\section{Appendix B: Demonstration of the Occurrence of Completely Plastic Contact at Very Small Length Scales}

In Appendix C of Persson's Journal of Chemical Physics paper on contact mechanics\cite{persson3}, he obtains a differential equation for the area of contact due to plastic deformation alone by integrating his diffusion equation:
$${\partial P(\sigma,\zeta)\over \partial\zeta}=f(\zeta){\partial^2 P(\sigma,\zeta)\over\partial\sigma^2} \eqno (1B)$$
over $\sigma$, to obtain 
$${\partial\over\partial\zeta}\int^{\sigma_Y}_0 d\sigma P(\sigma,\zeta)=f(\zeta)[-{\partial P(\sigma,\zeta)\over\partial\sigma}|_{\sigma=0}+{\partial P(\sigma,\zeta)\over\partial\sigma}|_{\sigma=\sigma_Y}] \eqno(2B).$$
The integral $\int^{\sigma_Y}_0 d\sigma P(\sigma,\zeta)$ is by definition $P(\zeta)=A_{el}(\zeta)/A$. The first term on the right hand side of Eq. (2B) was shown in Ref. \cite{persson3} to be equal to $-dA_{non} (\zeta)/d\zeta$, the negative of the derivative of the surface area not in contact with respect to $\zeta$. In the limit as $\sigma_Y$ approaches infinity, the second term on the right hand side of Eq. (2B) vanishes. Since in this limit there is clearly only elastic deformation, the latter term must be associated with plastic deformation. Then in this limit, Eq. (2B) can be written as 
$${dA_{el}\over d\zeta}+{dA_{non}\over d\zeta}=0 \eqno(3B)$$
and hence $A_{el}+A_{non}$ must equal a constant $A$ in this limit, and that constant must be A. Then, clearly when $\sigma_Y$ is finite, we must identify the last term on the right hand side of Eq. (2B) with $dA_{pl}/d\zeta$. Using his solution for $P(\sigma,\zeta)$ Persson finds by integrating this expression that 
$$P_{pl}(\zeta)={A_{pl}(\zeta)\over A}=$$
$$-(2/\pi)\sum_{n=1}^{\infty} (-1)^n ({sin\alpha_n\over n}\{1-exp[-\alpha_n^2\int_1^{\zeta} d\zeta' g(\zeta')]\}, \eqno (4B)$$
where $\alpha_n=n\pi\sigma_0/\sigma_Y$, and
$$P_{non}(\zeta)={A_{pl}(\zeta)\over A}=$$
$$(2/\pi)\sum_{n=1}^{\infty} ({sin\alpha_n\over n})\{1-exp[-\alpha_n^2\}$$
$$P_{pl}(\zeta)={A_{pl}(\zeta)\over A}=$$
$$-(2/\pi)\sum_{n=1}^{\infty} (-1)^n ({sin\alpha_n\over n})\{1-exp[-\alpha_n^2\int_1^{\zeta} d\zeta' g(\zeta')]\}. \eqno (5B)$$ 
Since as $\zeta$ approaches infinity $\int_1^{\zeta} d\zeta' g(\zeta')$ approaches infinity. we find that in the infinite $\zeta$ limit, 
the sum of Eqs. (4B) and (5B) becomes
$$P_{non}+P_{pl}={A_{non}(\zeta)\over A}+{A_{pl}(\zeta)\over A}={4\over\pi}\sum_{n=1,odd}^{\infty} {sin({n\pi\sigma_0\over \sigma_Y})\over n}=1, \eqno (6B)$$ 
by a well known identity, which implies that none of the asperities at the regions of contact for $\zeta=\infty$ are in elastic contact. 
Since there is no stress acting on the area $A_{non}$ by definition and $\sigma=\sigma_Y$ at all areas of plastic contact, we must have
$$\sigma_Y A_{pl}=\sigma_0 A. \eqno (7B)$$
This result also follows if we take the $\zeta$ approaches infinity limit in Eq. (4B), making use of the identity
$$-x={2\over\pi}\sum_n (-1)^n sin(n\pi x), \eqno (8B)$$
for $-1<x<1$. 

\section{Appendix C: A Demonstration that Archard's Model of Multiscale Asperities Predicts that the Contact Area Per Asperity Increases with Increasing Load}

Archard\cite{archard} presented a model for elastic distortion of multiscale asperities in which at the longest length scale an asperity has a hemispheric shape and distorts according to Hertz theory\cite{johnson}. The area of contact of such an asperity with a flat surface is assumed to be covered by a continuous distribution of smaller scale hemisherically shaped asperities. The largest length scale asperity described above is only in contact at the areas of contact of these smaller asperities. The smaller asperities are in turn also assumed to be in contact only at the locations  of a continuous distribution of still smaller asperities. Archard argues that as we consider shorter and shorter length scales, the contact area becomes more and more nearly proportional to the load carried by the original longest length scale asperity. 

At his longest length scale, Archard finds that the contact area is proportional to $W^{2/3}$, where W is the load and the asperity density is independent of load. At the next smaller length scale considered by him, the contact area is proportional to $W^{8/9}$ and the number density of these asperities is proportional to $W^{2/3}$, and hence the contact area per asperity (i.e., the quotient of these two) is proportional to $W^{2/9}$. At the next smaller length scale, the contact area is proportional to $W^{26/27}$ and the asperity density is proportional to $W^{8/9}$, leading to a contact area per asperity (the quotient of the contact area and asperity density) of $W^{2/27}$. Next order gives a contact area proportional to $W^{44/45}$ with an asperity distribution proportional to $W^{26/27}$, leading to a contact area per asperity of $W^{2/135}$. As we continue this process, the asperity density is always proportional to the same power of W as the contact area at the previous length scale. Hence, it is clear that as we go to successively smaller length scales, the contact area per asperity is proportional to a smaller and smaller power of W. Thus, it is clear that as W increases, the smaller length scale asperities' contact area will grow, suggesting that the smallest length scale asperities will flatten out as the load increases. Since the amount that asperities on smaller and smaller length scales flatten out as a function of W decreases as we go to smaller and smaller length scales. Since this theory does not assign a precise radius to each length scale asperity, it is difficult to assess the degree to which smaller length scale asperities flatten out (i.e., how the contact area per asperity compares to the original size of an asperity. Since the contact area at each length scale appears to be proportional to the inverse asperity density raised to the same power as the power of W that the contact area is proportional to at that length scale, we can take that quantity as related to the original asperity size. Since the contact area per asperity grows as W to a smaller power, it is possible that the asperities do not flatten out in Archard's model, in the sense that the contact area remains small compared to the asperity size as W increases.

\end{document}